\documentclass[aps,epsfig,float,twocolumn,pre,showpacs]{revtex4}

\usepackage[pdftex]{graphicx}
\usepackage{epstopdf}
\newcommand{\ignore}[1]{}

\begin{document}

\title{Human dynamics revealed through Web analytics}

\author{Bruno Gon\c calves}\email{bgoncalves@physics.emory.edu} 
\affiliation{Physics Department, Emory University, Atlanta, Ga 30033}

\author{Jos\'e J. Ramasco}\email{jramasco@isi.it}
\affiliation{Complex Systems Lagrange Laboratory, 
Complex Networks (CNLL), ISI Foundation, Viale S. Severo 65, I-10133 Turin, 
Italy}

\date{\today}

\begin{abstract}
When the World Wide Web was first conceived as a way to facilitate the 
sharing of scientific
information at the CERN (European Center for
Nuclear Research) few could have imagined the role it would come to play in the
following decades. Since then, the increasing ubiquity 
of Internet access and the frequency with which people interact 
with it raise the possibility of using the Web to better observe, 
understand, and monitor several aspects of human social  
behavior. Web sites with large numbers of frequently returning 
users are ideal for this task. If these sites belong to companies or 
universities, their usage patterns can furnish information about the working 
habits of entire populations. In this work, we analyze the properly 
anonymized logs detailing 
the access history to Emory University's Web site. Emory is a medium size 
university located in Atlanta, Georgia. We find interesting structure in 
the activity patterns of the domain and study in a systematic way the main 
forces behind the dynamics of 
the traffic. In particular, we find that linear preferential linking, 
priority based queuing and the decay 
of the interest for the contents of the pages are the
essential ingredients to understand the way users navigate the Web. 
\end{abstract}


\maketitle

\section{Introduction}

The access to Internet has become increasingly popular during the last decade.
However, despite
its importance, much is still unknown about the Web intrinsic properties,
the way people interact with it, and how it impacts our culture 
\cite{berners-lee06,watts07,romu04,sergei03}. 
Several theoretical approaches have been proposed in the last few years 
\cite{watts98,huberman98,barabasi99,menczer04,sergei01,ciro06,simkin07,wu07} 
but some fundamental 
issues remain yet to be fully understood. In this work, we will focus  
on answering the following question. 
Do any laws govern the way and frequency with 
which a person visits a given Web site or is each individual intrinsically
unique? From a sociological point of view, we 
would expect that, although the
 behavior of a single individual is ultimately personal and unpredictable, many
inferences can be obtained about the most common behaviors 
\cite{watts07,borgatta00}. A better understanding
of the way an individual uses a given Web site has
important economic consequences, as it can help the developers of
the site optimize it in a way that facilitates its use, and 
monetization. Apart from the utilitarian point of view, the activity patterns 
on the sites provide also important information on the dynamics of a 
population. The interaction with electronic devices or virtual 
instruments, such as social sites or mobile phones,
opens promising research avenues in this direction
\cite{onnela07,golder06,candia07,vazquez07,meiss08,zhou1,zhou2}.

\begin{figure}
\begin{center}
\includegraphics[width=8.cm]{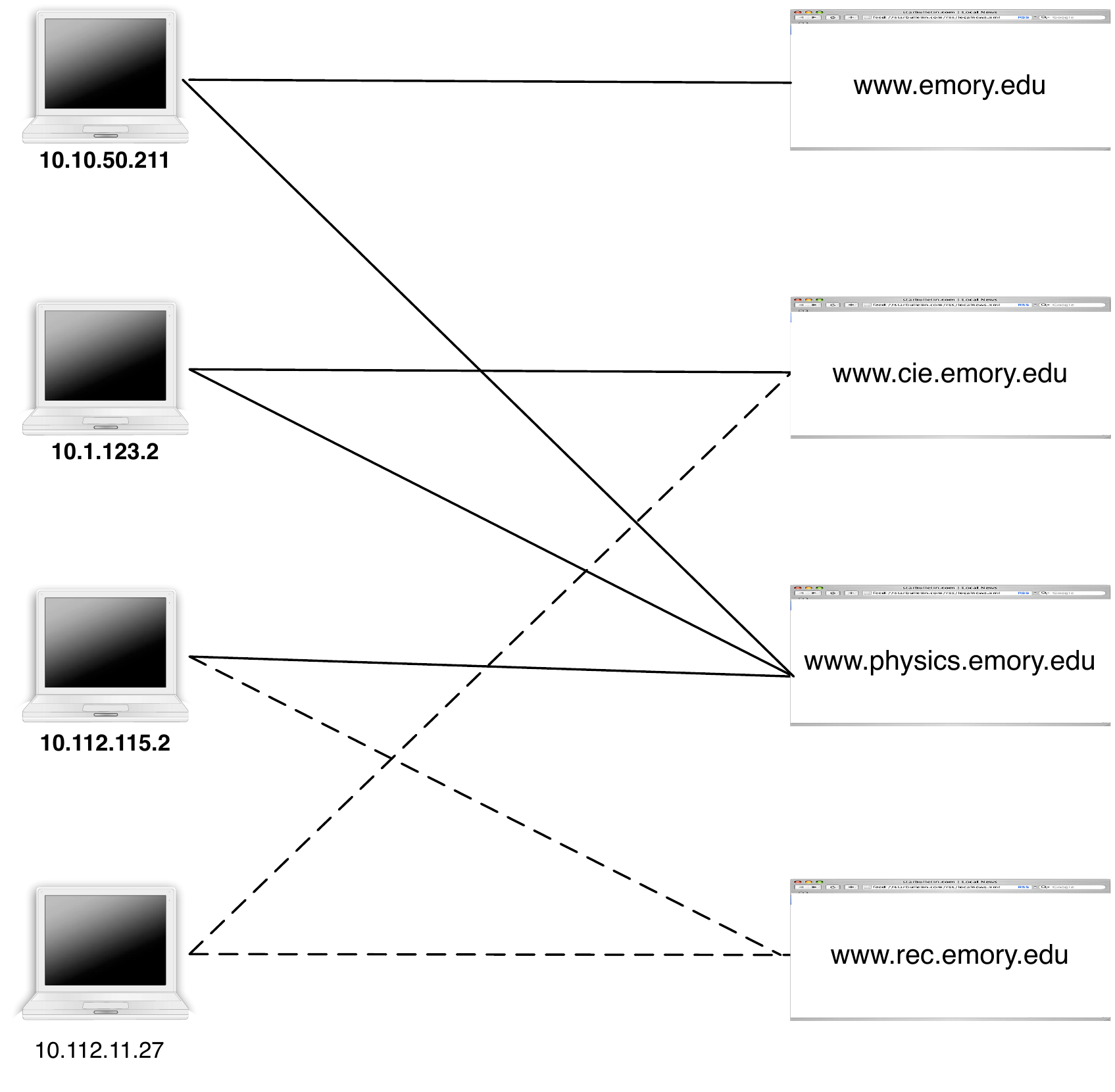}
\caption{Schematic representation of the interactions between 
users and Web pages. The system is dynamic, to provide a more visual impression
of its variability dashed lines represent new added connections.
\label{fig:diagram}}
\end{center}
\end{figure}

The sheer size and diversity of the 
World-Wide Web renders the 
attempts to characterize it on a global scale hardly feasible. Still 
several works have recently centered in describing from a statistical 
perspective the structure of the Web \cite{albert99,dill02}. If instead of 
understanding its structure the goal is to track how
users navigate it, the challenge becomes even greater. A  
solution consists
in ignoring the identity of the users, focusing only on the number of visitors
per site and on the number of clicks on its 
hyperlinks \cite{meiss08,meiss05}. Another 
possibility is to concentrate the attention onto a group of 
volunteers \cite{catledge95} or onto the users of a
social site that are      
usually well identified \cite{ciro07,capocci06,vazquez06}.
Our aim here is to follow the activity of individually trackable Web surfers in a relatively open 
environment and
characterize the way in which the interaction between users and Web sites 
occurs. This 
is the reason 
why we analyze the logs of the Web server of Emory 
University. These logs registered the
requests by Internet users, internal or external, of Web pages in the second level of the Emory 
domain (\emph{www.emory.edu}). The 
data comprehends a period that goes from
 Apr. $1$, $2005$ to Jan. $17$, $2006$. Each 
time a computer connects to the Internet, it is assigned a 
unique IP address that 
identifies it. 
When a user requests a page from a Web site, the IP, the page 
requested (URL), the time at which the request occurred and several other 
details are registered by the Web server. In our case, to preserve 
privacy the data has been anonymized in a coherent way, allowing us 
to follow the behavior of each IP by a single ID number but masking the real
identity. The log structure 
is represented schematically 
in Fig.~\ref{fig:diagram}.
On the left, we have the anonymized IP addresses which connect to the URLs 
on the right. To avoid the consideration of different elements of a Web 
such as photos or logos as independent pages, we have restricted our definition of URL to 
{\it (s)htm(l)}, {\it cfm}, {\it php}, {\it asp(x)}, {\it jsp} and {\it txt}
 documents. Each 
line of the logs corresponds to a different connection, that is 
timestamped with the 
date and time at which it took place. During our observation period, the domain 
received over $3$ million visitors to about $2.5$ million pages for a 
grand total of over $53$ million clicks.

\section{Activity Patterns of the Population}

\begin{figure}
\begin{center}
\includegraphics[width=8.cm]{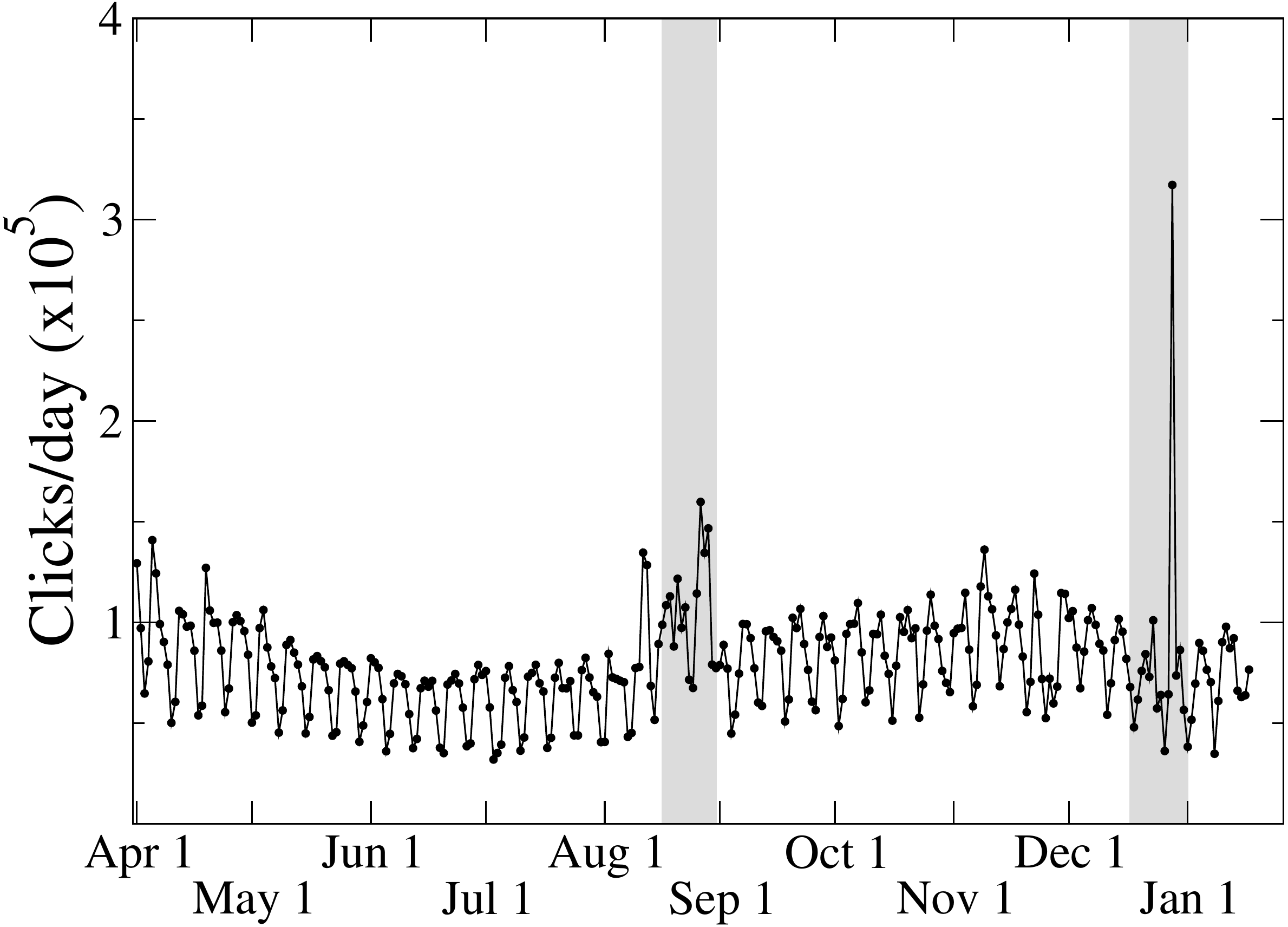}
\caption{Total number of clicks registered per day during the whole period 
of traffic observation. The gray bands correspond to the beginning and the end
of the semester: from Aug $16$ to Aug $31$, and  from Dec 
$16$ to Dec $31$.\label{fig:traffic}}
\end{center}
\end{figure}

Let us start by taking a view of the collective behavior of the entire 
population during the time period for which we have data. Intuitively, we 
expect the activity on a domain to vary from day to day, week to week and 
even 
month to month. 
In particular, it should be possible to observe variations in the activity, 
seen as the number of requests, due to weekends, holidays and other major 
events that disrupt the normal life of the University. The 
traffic at Emory is dominated by students and professors in the course of their 
professional activities 
and hence the major events in the course of the school year, 
such as the beginning and end of a semester, breaks or holidays, should be 
noticeable in the Web traffic. 
In order to check this idea, the number of page requests detected per day is 
shown in Fig.~\ref{fig:traffic} as a function of the observation date. One obvious 
feature of the figure is a clear oscillatory behavior with a period of one
 week. It 
also displays different trends for two special times of the year: one at
the later part of August, corresponding to  
the beginning of the school year, and the other at the end of December, when 
the semester finishes.

\begin{figure}[b]
\begin{center}
\includegraphics[width=8.cm]{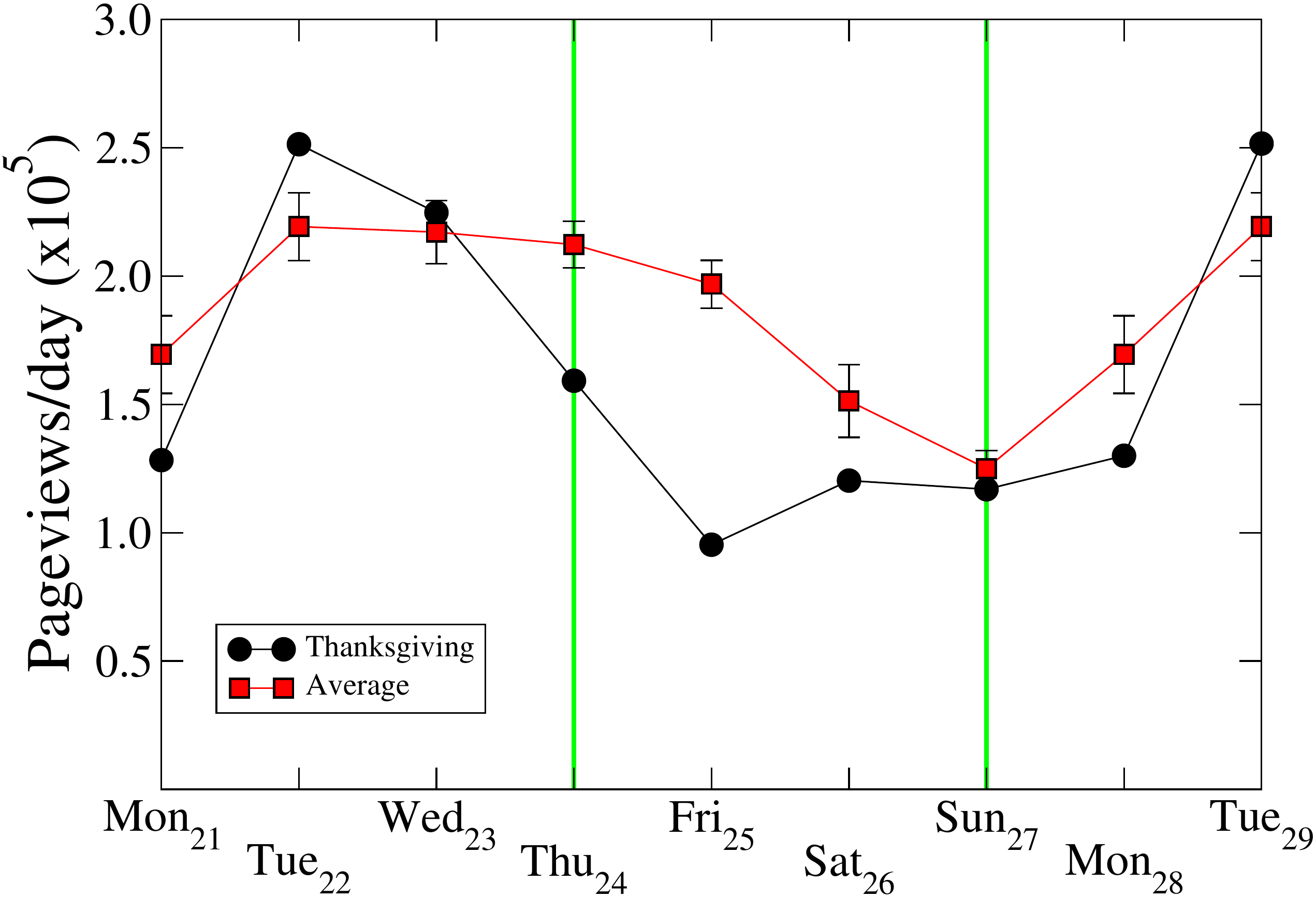}
\caption{Comparison between the average week activity and 
activity during Thanksgiving week. The green vertical lines represent the 
beginning and end of the official Thanksgiving break at Emory University. The
error bars for the average are calculated as two times the standard deviation, 
$2 \sigma$, or the $95 \%$ confidence interval.\label{fig:thanks}}
\end{center}
\end{figure} 

Since accesses to Emory domain are mostly work related, traffic can be used as
an indirect measure of the University "productivity".
Busier days would result in larger amounts of traffic, while during 
holidays and weekends the number of page requests is overall smaller, thus 
rendering
the relative changes in traffic significant. 
The 
averages of page requests by day of the week during the complete 
observation period are plotted, 
together with their corresponding $95 \%$ confidence intervals, in 
Fig.~\ref{fig:thanks}. 
Our results 
support the old adage 
that after Wednesday, the hardest part of the week is already behind us, with 
the activity slowly decreasing from then on to the weekend. Sundays are 
the least active day of the 
week. It is also interesting to note the not-so active behavior of Mondays, only
slightly more active than Saturdays. 
Armed with an estimate of how 
activity evolves over the week, we are now in a position to evaluate the 
effects
of a break. 
In the same Figure, we also represent the 
data for the days surrounding Thanksgiving, one of the major holidays 
in 
the US. Traditionally, Thanksgiving recess goes from 
Thanksgiving 
Thursday till Sunday, so one might expect any decreases in activity to be most
 noticeable during this period. 
This is 
what we observe, but we find other 
effects as well. Both the Monday before and after 
Thanksgiving seem to be less productive than normal. This is however 
complemented with busier than usual Tuesdays before 
and after the break.

\begin{figure}
\begin{center}
\includegraphics[width=8.cm]{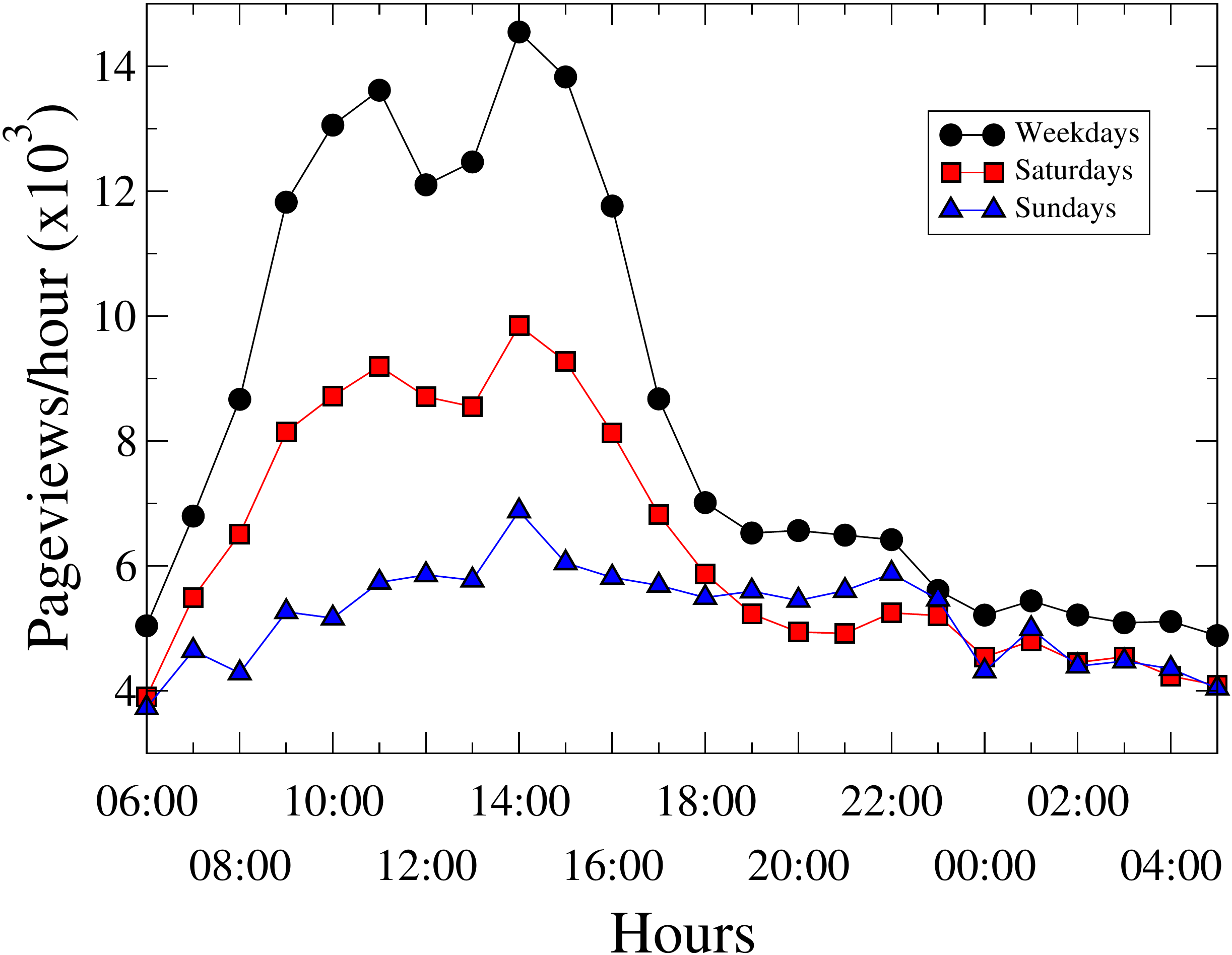}
\caption{Average hourly activity in the complete 
Emory domain as a function of the hour of the day. The curves are averaged over
the weekdays (circles), Saturdays (squares) and Sundays (triangles).
\label{fig:hours}}
\end{center}
\end{figure}

Intra day variations, with some times of the day being busier than others are 
also 
seen. By averaging the activity observed at a given hour over all the weekdays 
in our data set, we obtain Fig.~\ref{fig:hours}. The most active period 
is 
between $7$AM and $6$PM. The large dip between 
$11$AM and $2$PM is due to the lunch 
break. After lunch, the activity peaks reaching the higher level 
of the day. After $6$PM activity levels off until $10$PM, 
marking the end of the workday. Saturdays do not differ significantly from
other days of the week, only Sundays display a different activity profile.  
Similar patterns for human circadian rhythms have been recently 
reported for other systems in Refs. \cite{golder06,vazquez07,meiss08}. 
Such ubiquity indicates important universal features (profiles) 
regarding human habits that Web analytics can help to characterize in a
quantitative way.

\section{Individual Activities} 

\begin{figure}[b]
\begin{center}
\includegraphics[width=8.cm]{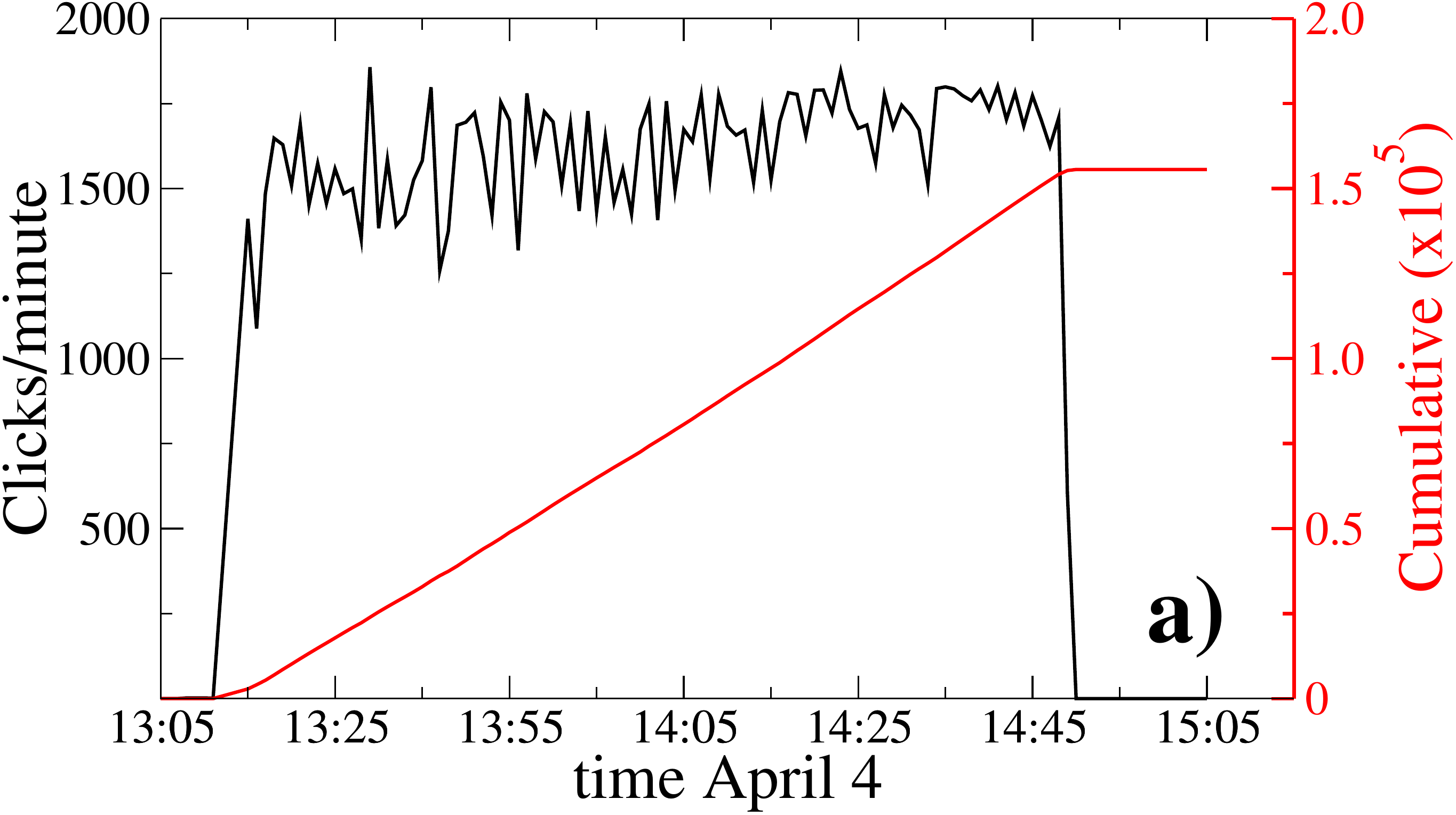}
\includegraphics[width=8.cm]{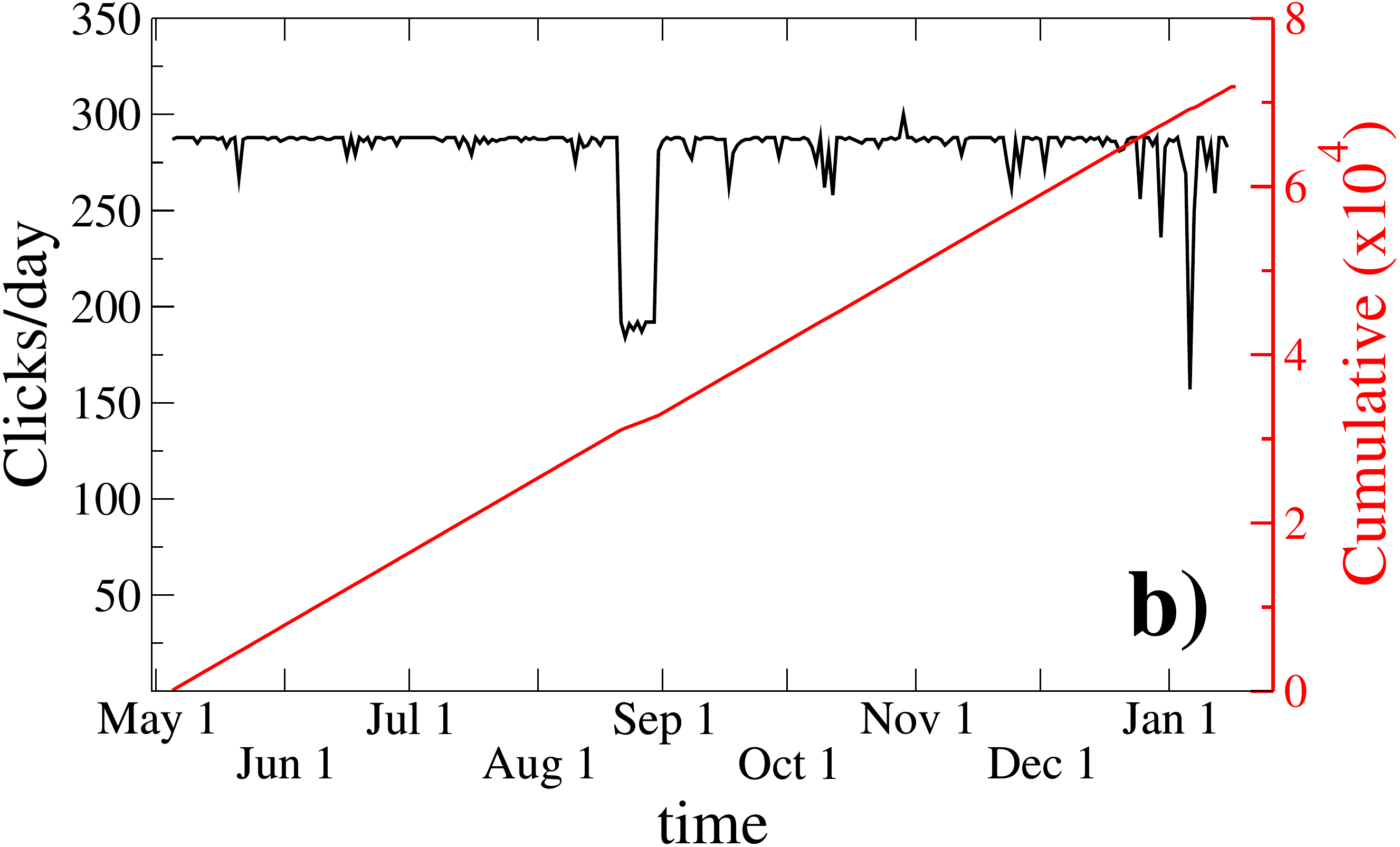}
\includegraphics[width=8.cm]{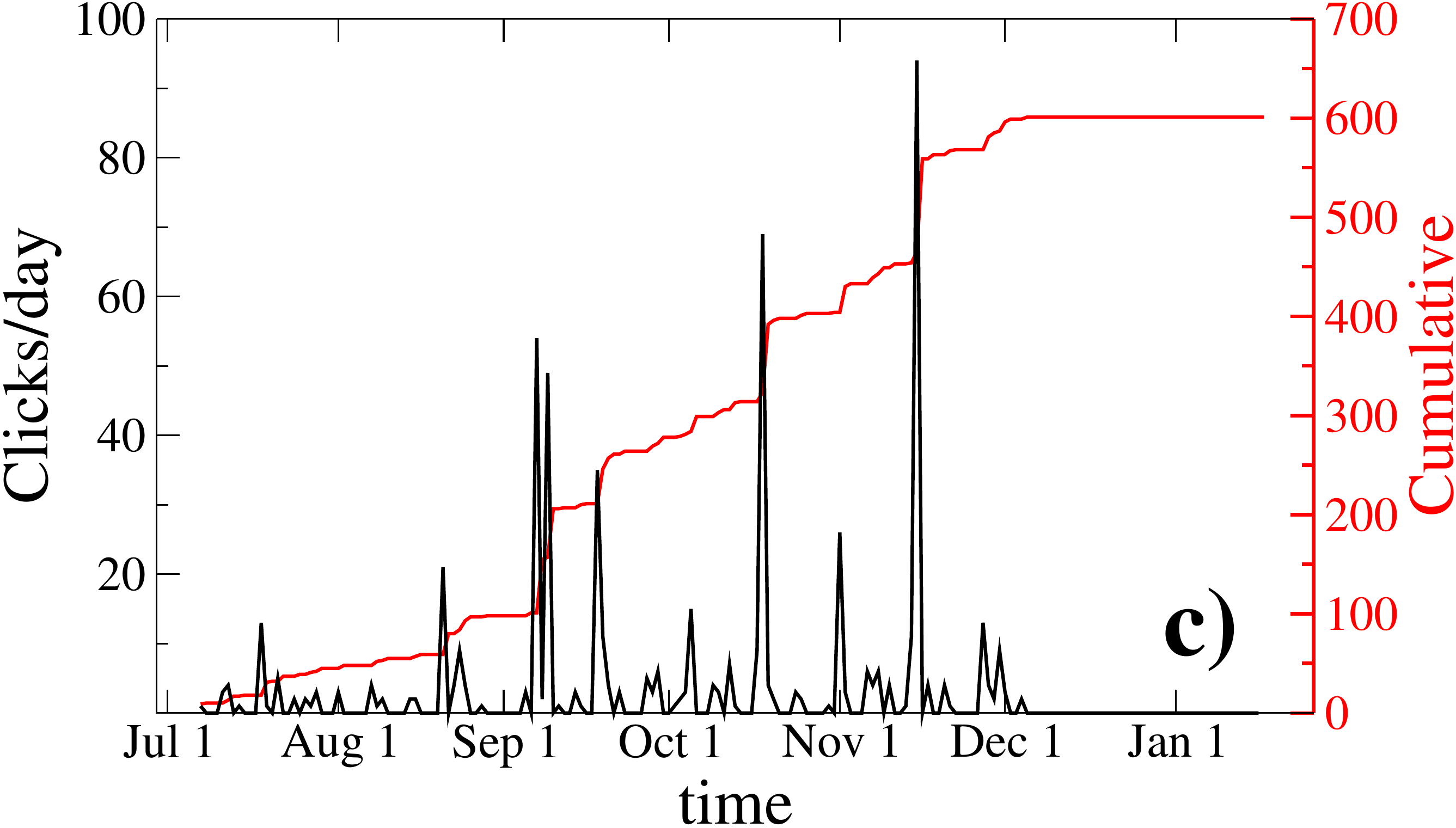}
\caption{Activity history of several individuals: a) what 
seems to be a malicious attack on a finance Web page of the University, b) an 
automatic software update program, and c) a human user filling data in an 
administration site. The red curves represent the cumulative number of 
clicks. To facilitate the visualization, the scale of the cumulative and 
temporal number of clicks are different. The axis on the right side of each 
plot displays the scale for the cumulative number of clicks. 
\label{fig:hist}}
\end{center}
\end{figure}

Although interesting, the analysis of averages taken over the entire population 
has limitations. The histograms of single user activity are 
typically very wide, being in 
some cases well-modeled by power-law distributions with exponents smaller than 
$2$ \cite{meiss05}. When this happens, it is difficult to identify a 
"typical" user based on such metrics: while 
most users only visit the 
domain sites a few times, a significant fraction of individuals (as identified 
by their IP addresses) accumulate large numbers of page requests. 
This 
variability deserves greater attention since it can carry important 
information. Figure~\ref{fig:hist} shows the activity patterns 
of three users. We do not know the actual IPs 
but it is possible to deduce the intention of the visit based on the 
particular URL accessed and on the profile of the activity. In 
Figs.~\ref{fig:hist}$a$ and \ref{fig:hist}$b$, the users are computer programs. 
One, the case shown in $a)$, corresponds to a malicious attack on an finance 
service Web page
of Emory. It took place on April the $4$th. The profile of the number of access 
attempts per unit of time displays a very peculiar shape, quite regular as 
occurs for most automatic navigators, with a very high number of requests 
concentrated in a short period of time. Other, more friendly, robots 
are those corresponding to updating programs. An example can be seen in 
Figure~\ref{fig:hist}$b$ where a software site in Emory is regularly visited 
presumably in search for new updates. Finally, human users show a very 
different activity profile from that of the machines. The activity of a human 
user selected at random can be seen in Figure~\ref{fig:hist}$c$. In this case 
the URL is an administrative site that demands manual introduction of data. 
The activity congregates in some days followed by relative long periods of 
time without any request.

\begin{figure}
\begin{center}
\includegraphics[clip,width=8.cm]{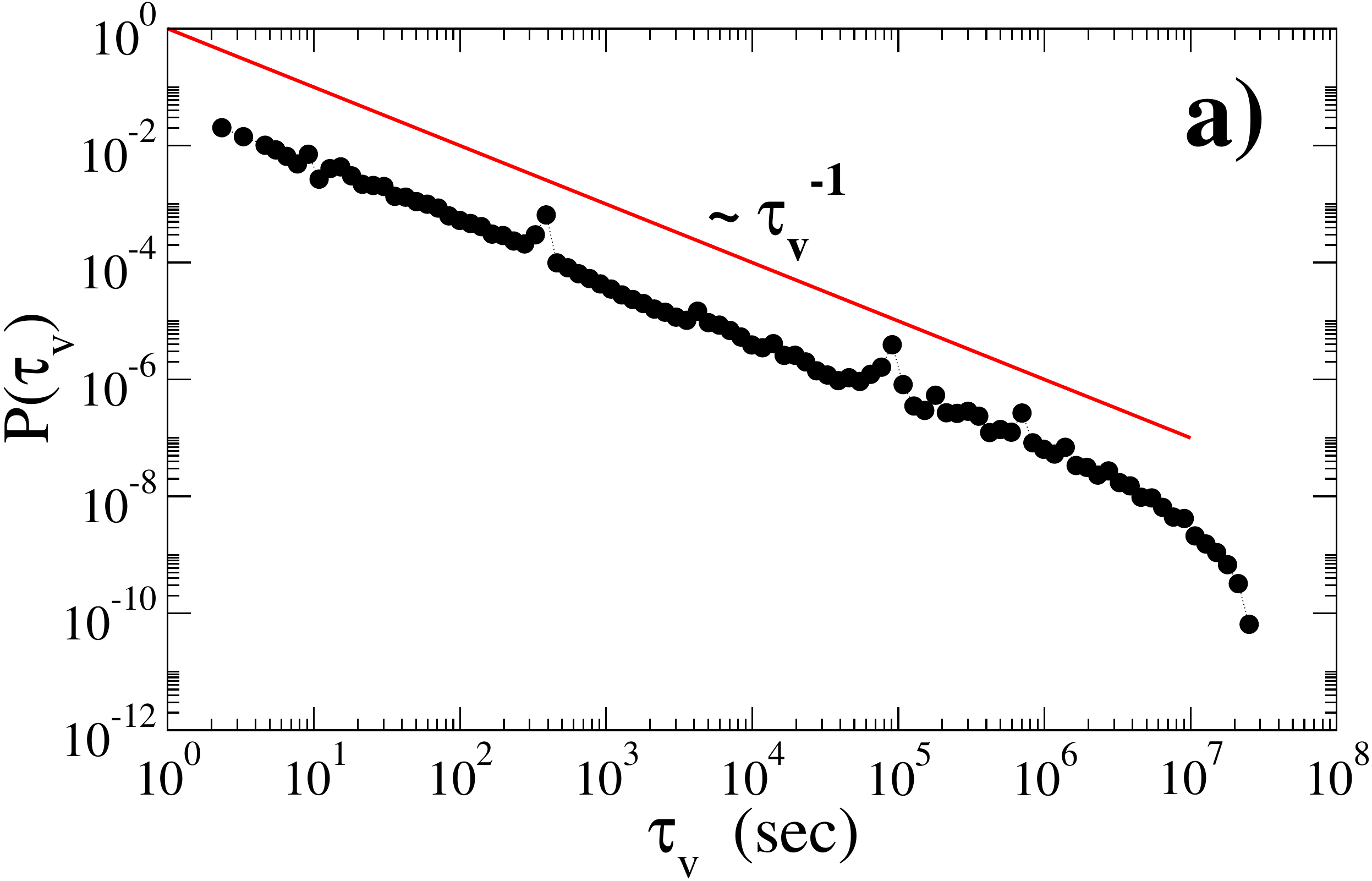}
\includegraphics[clip,width=8.cm]{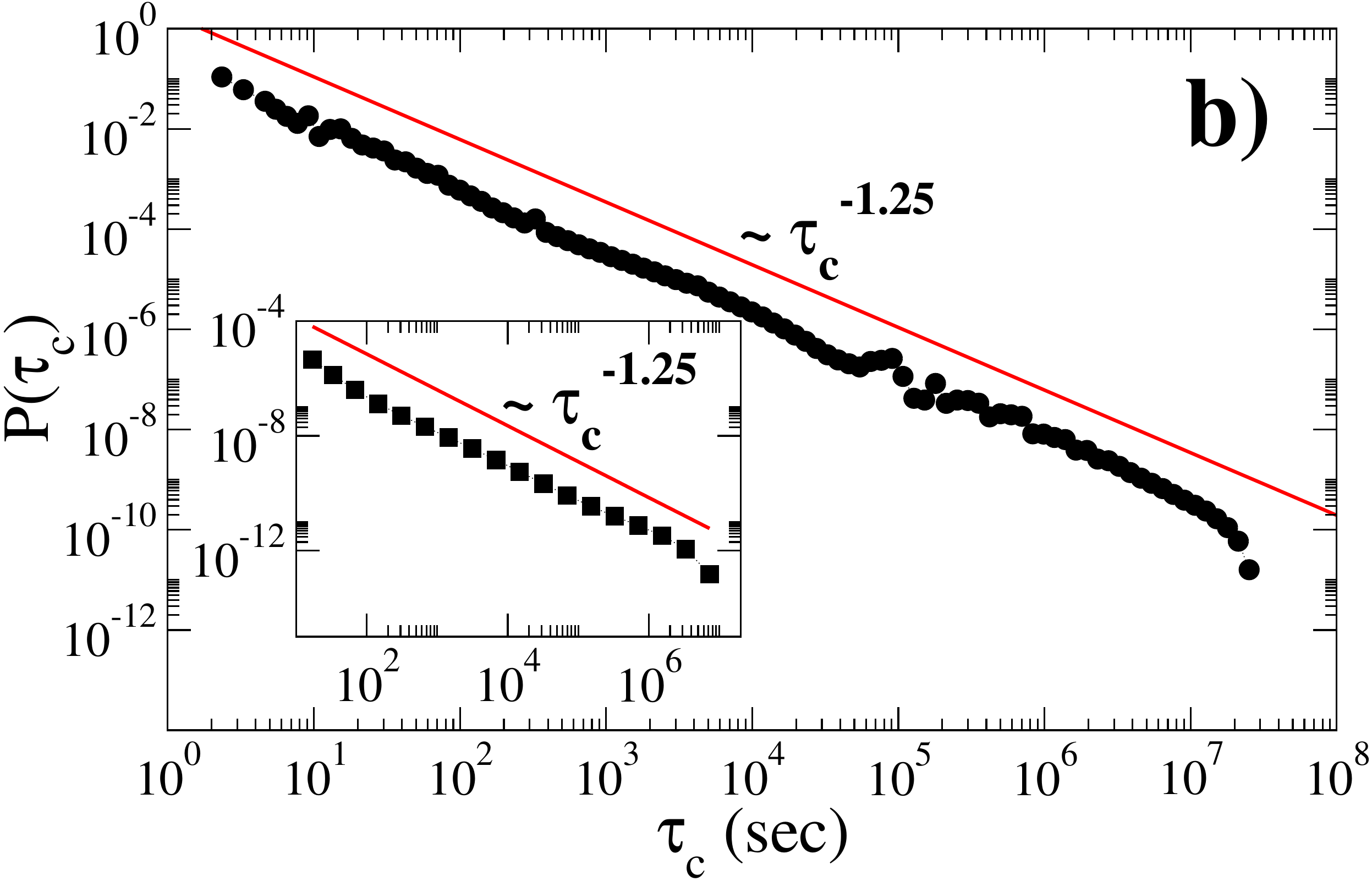}
\caption{\label{fig:ft} Distribution of times between consecutive clicks: a) 
visits of the same user to the same URL, and b) the same user to any page of 
the Emory domain. The straight lines correspond to the power-law 
$f(\tau) \sim \tau^{-1}$ in a) and to $f(\tau) \sim \tau^{-1.25}$ in b). In 
the inset of b), the distribution of time in the queue is plotted for a 
variation of Barab\'asi's model \cite{barabasi05} (see text) with a number of executed tasks per unit of 
time of $\nu = 3$, with probability of choosing a task according to priority 
$p = 0.99999$, a total of $100$ tasks and $10^7$ time 
steps.}
\end{center}
\end{figure}

Given the strong variability in the activity of human users, it is interesting 
to measure some statistics about it. In Figure~\ref{fig:ft}, we have represented 
the histograms of the duration of the periods between requests for two different 
scenarios: in Fig. \ref{fig:ft}$a$ for the time between consecutive visits of 
the same user to the same URL, $P(\tau_v)$, and, in Fig.~\ref{fig:ft}$b$, 
for the time between clicks by the same user to any of the sites in Emory's 
domain (not necessarily to the same URL), $P(\tau_c)$. Both distributions 
are rather wide. 
The distribution $P(\tau_c)$ can be well fitted by 
a power-law decaying function of the type $P(\tau_c) \sim \tau_c^{-1.25}$. 
The distribution of time between consecutive visits, $P(\tau_v)$, decays 
even more slowly with an exponent of value $-1$. 
This latter value can be 
understood thanks to a model on human dynamics recently proposed by 
A.-L. Barab\'asi \cite{barabasi05} (see also \cite{vazquez07,vazquez06,
oliveira05,vazquez05,oliveira07}). 
In 
this model, an agent has to perform a set of tasks each with a random priority 
assigned. A step consists in the selection of the task with the highest 
priority with probability $p$ or of a random one with probability $1-p$. After 
the execution, a new tasks occupies the free spot in the queue. This group of 
rules is extremely simple but is able to reproduce a distribution of waiting 
times for the tasks in the queue that, in the limit of small $p$, decays as 
$\sim 1/\tau$. It can be argued that consecutive visits to the same site in
Emory are equivalent to one of these tasks since many of the visits are 
related to work or studies, and probably bear an inherent sense of priority 
for each user. Also returning immediately to the same URL and reloading it is 
not a common practice, at least not among humans. It is 
important to note that if the user pushes the back bottom in the browser,
typically we 
are not able to detect such a move because it does not leave a trace in the 
logs of the server due to browser caching. If each entrance is seen as a fresh 
start of a different task, 
the parallelism between the rules of the model and the way
users return to the same pages can be justified.

\begin{figure}
\begin{center}
\includegraphics[clip,width=8.cm]{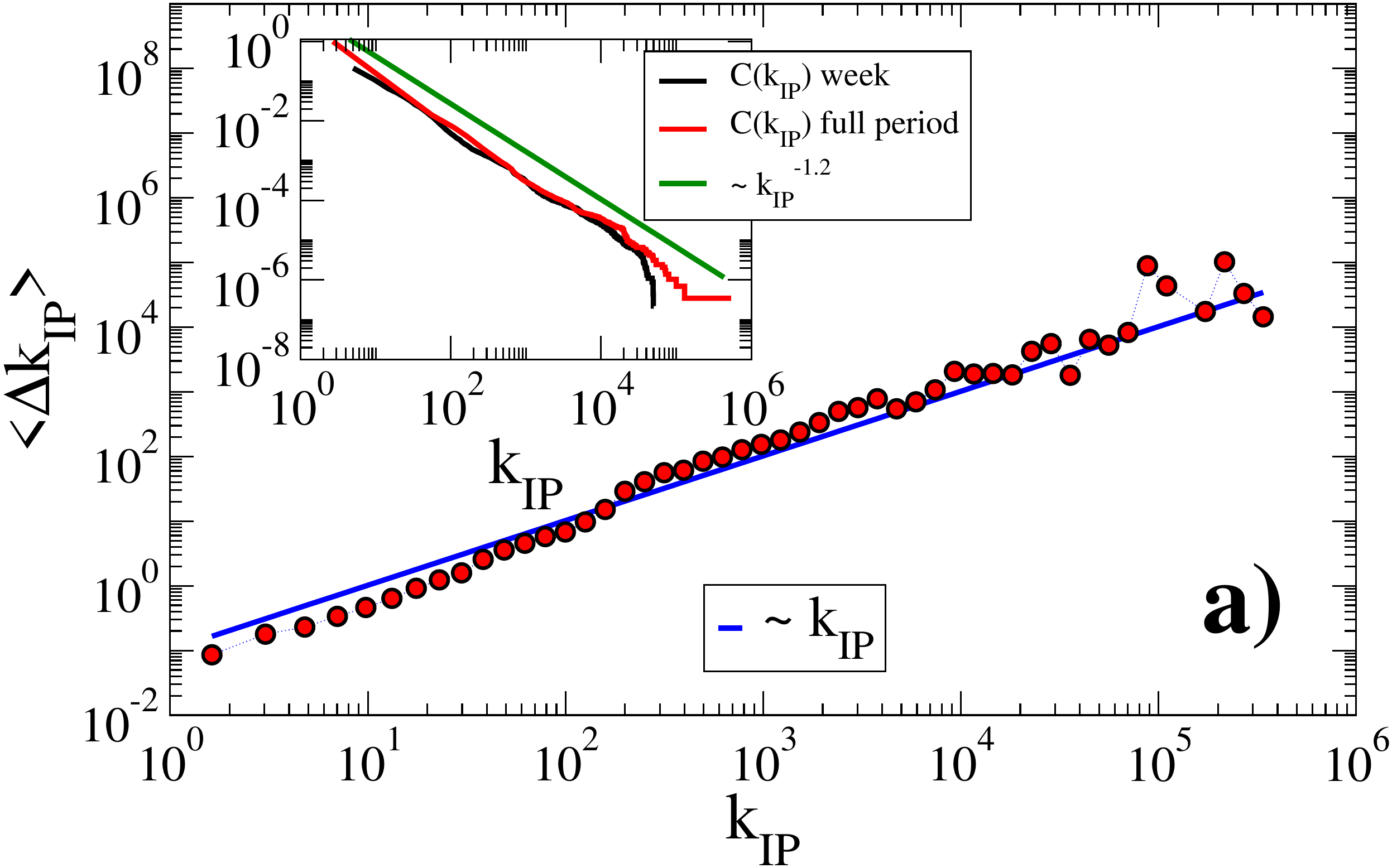}
\includegraphics[clip,width=8.cm]{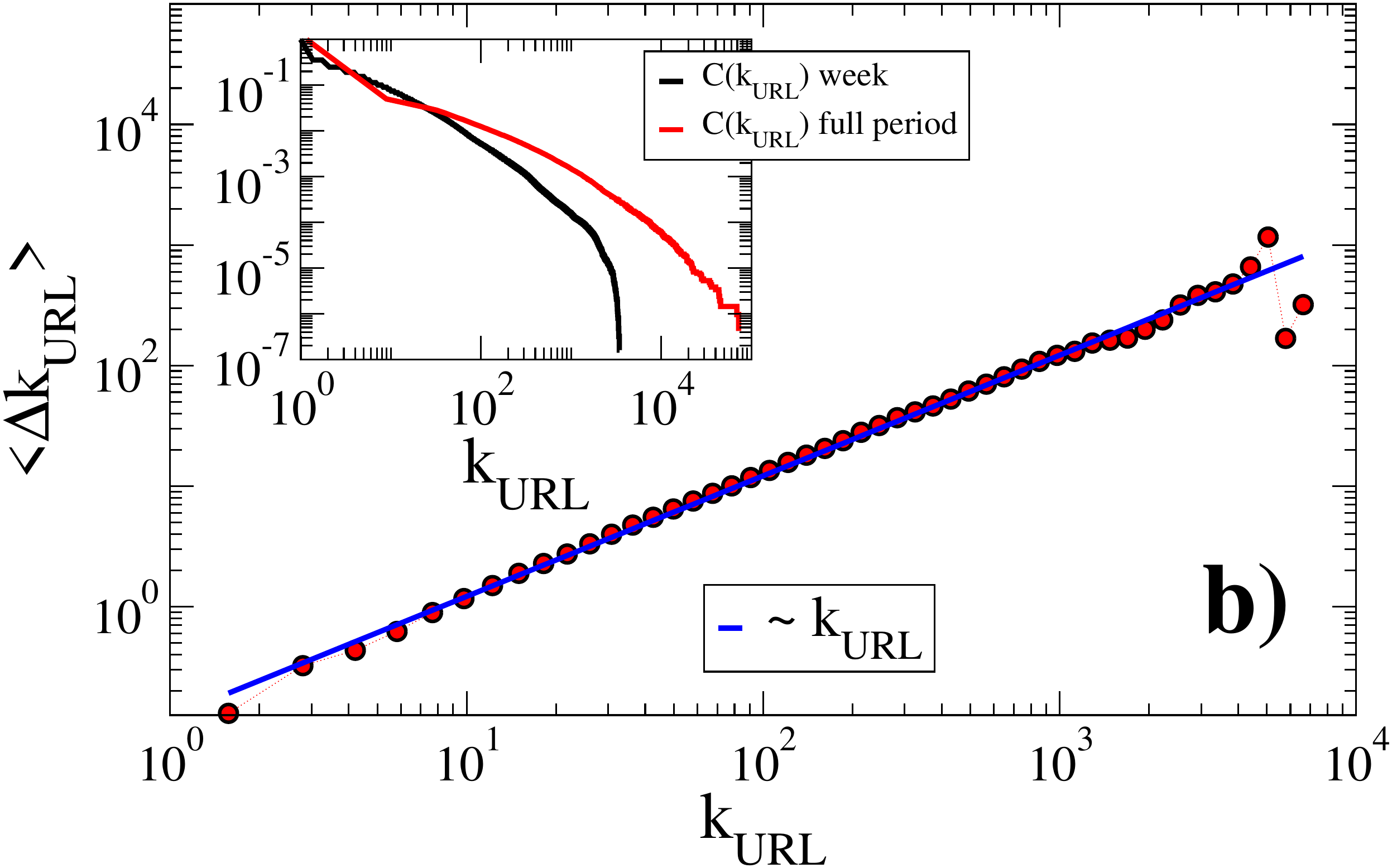}
\includegraphics[clip,width=8.cm]{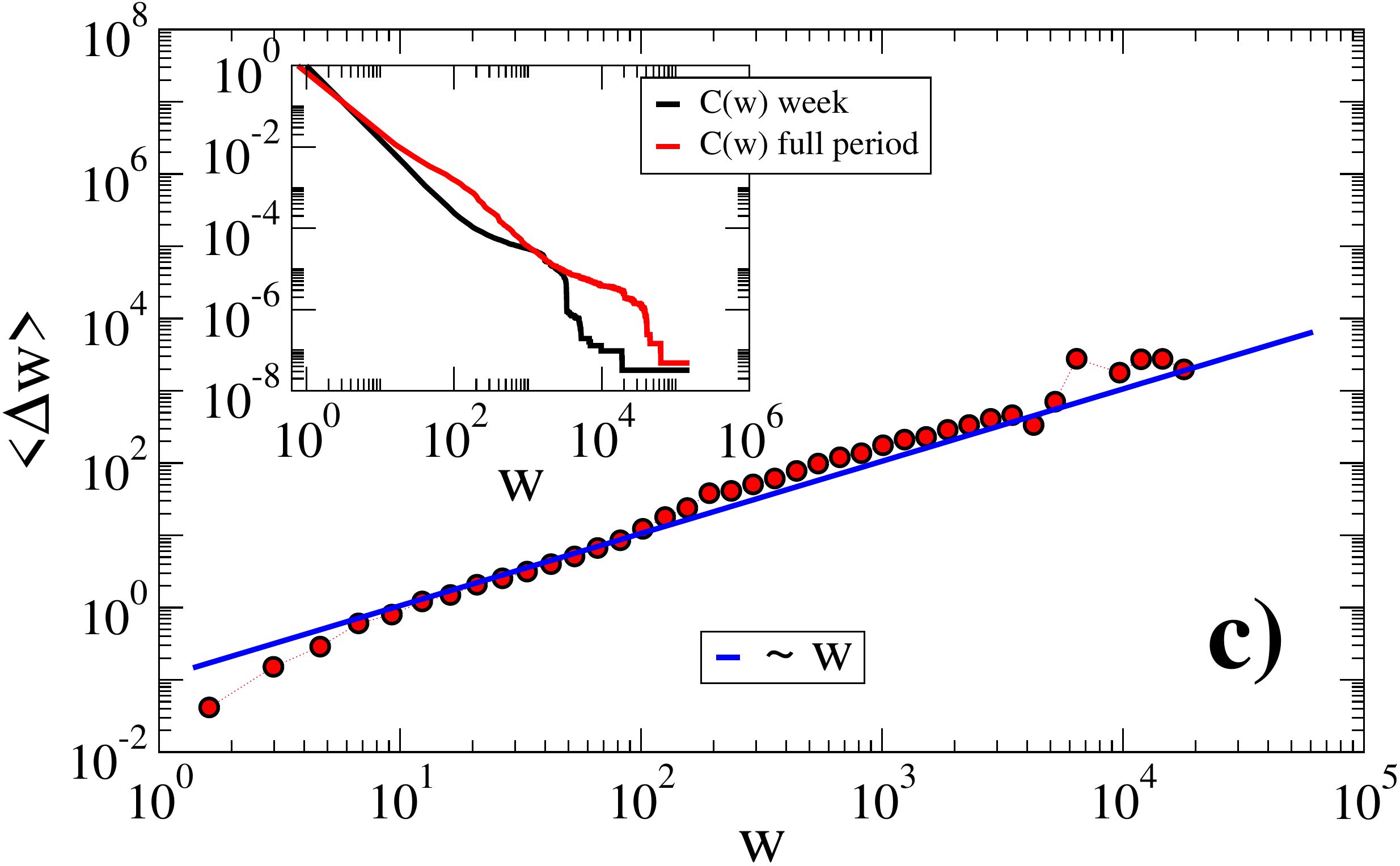}
\caption{\label{fig:dk} a) Average variation in a single day of the number 
of different visited sites, $\langle \Delta k_{IP}\rangle$ as a function of 
the number of sites already seen during the previous week, $k_{IP}$. b) The 
same type of function but for the number of visitors to an URL, 
$\langle \Delta k_{URL}\rangle$. And c) the average day variation of 
the number of clicks on each connection IP--URL as a function of the clicks 
accumulated during the previous week, $\langle \Delta w\rangle(w)$. The insets 
display the cumulative distributions for each quantity, the black curves are 
obtained by splitting the database in one week periods and average over all 
of them, while the red ones are the distributions for the full $292$ days 
period.}
\end{center}
\end{figure}

The question is then whether there is a way to understand also the exponent
$-1.25$ of $P(\tau_c)$. The answer is yes, if one considers that a single 
click on the domain does not necessarily have to be related to the realization 
of a task. Many tasks will require a (fast) sequence of clicks on different 
sites of the domain for their completion. This is why we propose the following
modification of the model: each time step, instead of a single task, a group of
$\nu$ tasks is selected for execution. The selection of each of them is 
done as before: by
priority with probability $p$, and at random otherwise. We have performed a
systematic numerical study of this model and found that provided that $\nu >2$
the exponent of the distribution of the 
time of permanence in the queue decays always 
as $\sim \tau^{-1.25}$. An example
with $\nu = 3$ is shown in the 
inset of Fig. \ref{fig:ft}$b$. These two models are 
oversimplifications but seem able to capture some of the essential features 
present in the dynamics of a large community of users leading to the existence 
of universal exponents.

\section{Attractiveness and Preferential Linking}

Another aspect that is worth to explore in the dynamics of our database is 
whether the new connections or new clicks follow a preferential rule. 
Preferential linking or the "rich get richer" effect is a 
relatively old concept considered originally in a socio-economic context by
E.H. Simon \cite{simon55}. 
In the 
area of graphs theory, it was introduced in $1999$ \cite{barabasi99} with a
model inspired in the hyperlinks of the Web
(see also \cite{bornholdt01,dorogovtsev00}). A few years have 
passed, and although several attempts have been made to check the existence of 
preferential linking in a variety of systems \cite{capocci06,barabasi02,romu01,
redner05}, as far 
as we know, a systematic study of preferentiality on the user-Web relationship 
is still missing. To be precise, if the variable under consideration $x$ 
can change in time for each element of the system, it is said that it shows 
linear preferentiality if the variation follows on average an expression of the 
type
\begin{equation}
\langle \Delta x \rangle \approx A \,  x +B , 
\label{eqn:perf}
\end{equation}
where  the average $\langle .
\rangle$ is taken over all elements $i$ of the system with $x_i = x$, and $A$ and $B$ are 
constants. This mechanism supposes that if the update refers to quantities such 
as number of connections or number of clicks of a site, the probability that 
a particular site is chosen to update is proportional to the number of 
connections or clicks that it has previously accumulated. More popular 
sites concentrate thus higher attention leading to an agglomeration process 
that, after a while, produces a very wide distribution of values of $x$. If 
the relation of Eq. (\ref{eqn:perf}) is linear, the 
distribution $P(x)$ can be approached by a decaying power-law function with 
an exponent depending on the values of 
$A$ and $B$ \cite{sergei03}. If it is not linear, 
two simple scenarios can occur. Either $\Delta x$ grows with $x$ faster than 
linear and the most popular element will eventually congregate a finite 
fraction of all the available value of $x$, or 
it is sublinear and the distribution of values of $x$ will not be wide 
(stretched exponential instead of a power-law) 
\cite{sergei03,kraprivsky00,kraprivsky01}.

In our case, the "elements" of the system are Web pages and IPs, and the 
quantity $x$ can be, among other things, the number of clicks of a certain user 
on a given URL, which we call $w$, the number of different users that an URL 
receives $k_{URL}$ or the number of different sites that an IP visits 
$k_{IP}$. We have also performed a similar study for the activity of the 
URLs and IPs (defined as the number of requests received or sent), but the
results are similar. We will focus therefore our attention 
only on $k_{URL}$, $k_{IP}$ and $w$. The variation of each of these variables 
$\Delta x$ in a 
single day is measured after having accumulated 
the values of $x$ for a full week. Then an average is taken over all the weeks 
of the database. The results displaying $\Delta x$ as a function of $x$ are
depicted in 
Fig.~\ref{fig:dk}. The variation of $k_{IP}$, $k_{URL}$ and $w$ can be well
approached by linear preferential functions similar to Eq.~(\ref{eqn:perf}) 
(straight lines in the main plots). This means that the rate at which users
explore the Web ($\Delta k_{IP}$), the rate at which popular pages attract 
new users ($\Delta k_{URL}$) and the rate
at which users revisit Web pages ($\Delta w$) depend linearly on the previous week
performance. It should also imply that the distributions $P(k_{IP})$, $P(k_{URL})$
and $P(w)$ are wide and well fitted by a power-law. In order to check
this last point, we have measured the cumulative distributions $C(x) = \int_x^\infty
dy P(y)$ for the three quantities. The cumulative distribution $C(x)$ is the
probability of having a value of the variable greater than $x$ and usually 
exhibits better statistics than $P(x)$. Note that if 
$P(x)$ goes as $P(x) \sim x^{-\gamma}$, then $C(x) \sim x^{1-\gamma}$. The results 
are shown in the insets of Fig.~\ref{fig:dk}. In these plots, we have also
included the cumulative distributions estimated aggregating the values of $k_{URL}$, 
$k_{IP}$ and $w$ for the whole period of the database ($292$ days). The
comparison of the cumulative distributions obtained for the two time windows
reserves us an important surprise. For $C(k_{IP})$, the two curves overlap 
and can be fitted with a power-law of exponent $\gamma \approx 2.2$. However,
this is not true for the popularity of the URLs, $k_{URL}$, or for $w$.
This difference in the output depending on the extension of the time window 
has important 
consequences for modeling the dynamics of the system. Its origin is related to the
fact that in a university the time during which a site, or more specifically its 
content, is relevant closely 
tracks the evolution of the academic year. In 
general, a similar rule should apply to all the Web sites. The life time can 
be more flexible, depending also on the number of visitors, but a certain loss 
of  interest as the time passes since the first online publication can be
expected \cite{wu07}. After this time, the page does not attract new users or
visits from the old ones at
the same rhythm (if attracts any at all). This breaks one of the implicit
assumptions of preferential linking: new elements are added at a constant
rate, while the old ones keep attracting attention indefinitely. It also implies
that linear preferential linking is not valid for longer time windows for
$k_{URL}$ and $w$, and that their distributions cannot be modeled as simple
(stable over time windows) power-laws.

\begin{figure}[b]
\begin{center}
\includegraphics[clip,width=8.cm]{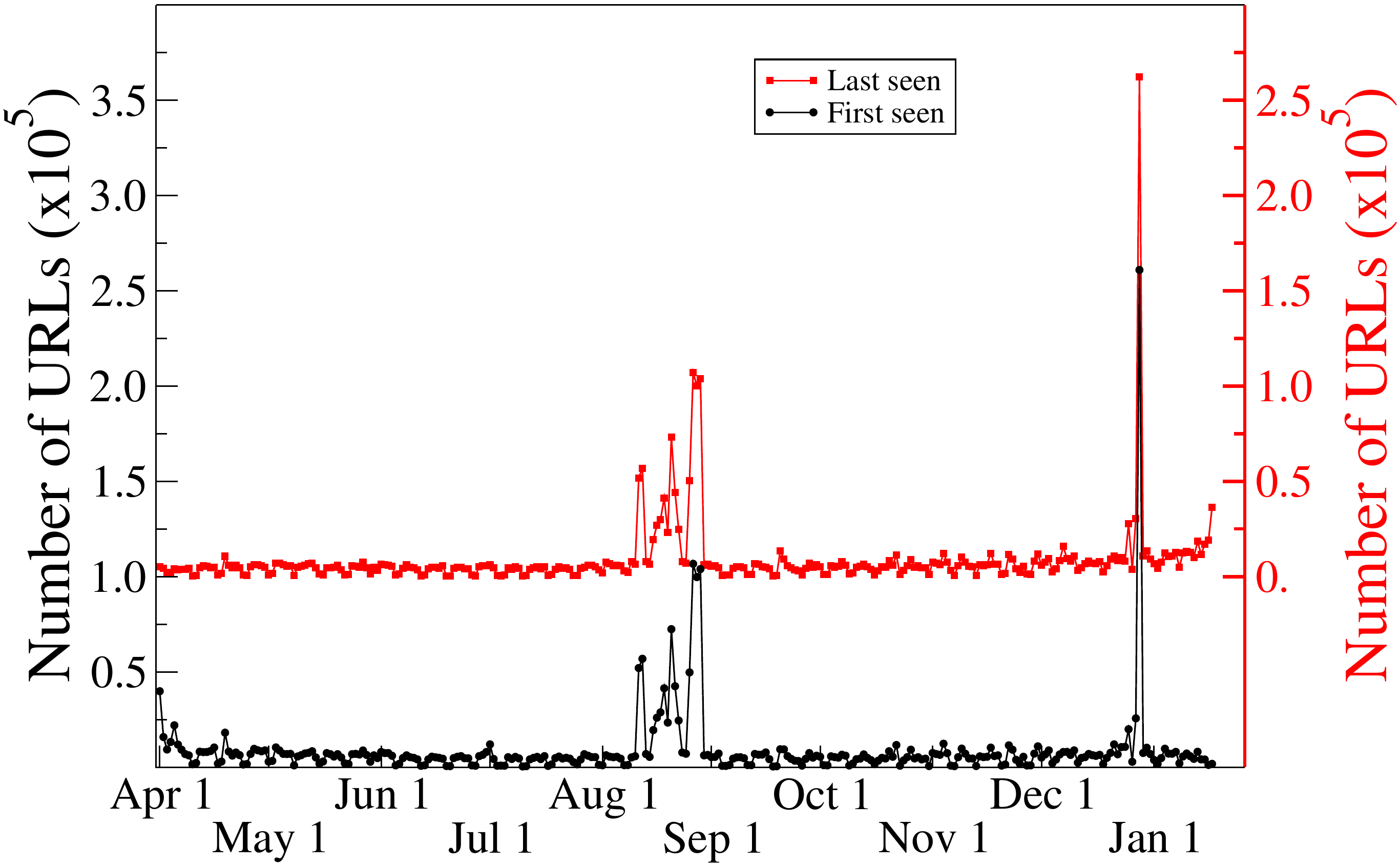}
\caption{\label{fig:present.URL}The number of URLs that are first (last) seen as
a function of time. The two major "extinction" and 
"creation" events, correspond to the beginning and end of the semester and
closely match the peaks detected in  
Fig.~\ref{fig:traffic}.}
\end{center}
\end{figure}

To visualize the life story of URLs, we represent in
Figure~\ref{fig:present.URL} the number of pages first seen or last seen in 
the system as a function of time.  We will say that a certain URL 
$\mathcal{U}$ is 
first seen at time $t$ if it receives its first request at $t$. 
Complementarity, the time in
which $\mathcal{U}$ is last seen,  
disappearing from the database, is when it receives the last registered visit.
Note that, although similar in look, this plot is different from 
Fig.~\ref{fig:traffic} where we are plotting the activity measured as the total
number of clicks on the Emory domain as a function of time. Two large peaks 
can be seen in Fig.~\ref{fig:present.URL}. The time of these peaks
coincides with the 
 end and beginning of the semester. Many Web pages seem thus to 
have a relative short life, probably being set up by professors or 
students that abandon them at the end of the semester. In many cases, even  
the \emph{http} addresses are no longer maintained.

\section{Discussion and Conclusions}

Web server logs have proven to be an important source of information
regarding human dynamics. Here we have offered an extensive study
on the medium size Web domain of Emory University 
tracking the users in a 
consistent way for $292$ days. 
A clear signal of human
circadian rhythms has been obtained as well as activity patterns that seem to be 
universal since they are in agreement with previous results on mobile phone 
records or email posting in social sites. In addition, 
in this case, the online traffic can be related to the productivity 
of the members of the University, namely, students, professors and
staff. The comparison between the activity of an ideal average
week and the week containing Thanksgiving is revealing in this sense, with 
some days concentrating an important level traffic, much higher than the 
average, and others falling clearly behind.

After the characterization of activity at the whole University scale, we have 
moved our focus down to the study of statistics of single users. The 
difference in the navigation
patterns between humans and automatic processes, either malicious or friendly,
has been highlighted. Humans are in general more unpredictable, although a
similar behavior might be reproduced by sophisticated automatic means. In 
particular 
for human users, it is important to analyze the statistics of the times 
between events (clicks) and compare them with recently introduced models based
on priority queues. We have shown that indeed such models are able
to explain the inter-clicks period distribution if the dyad user-site is
considered. Furthermore, a simple modification, in which the number
of tasks to execute in a short interval of time is higher than one, can also
account for the statistics of times between requests of the same user on the
whole Emory domain.

Finally, we have explored another mechanism that has been proposed as an
important ingredient in the development of the $WWW$, namely "preferential
attachment". Linear preferential attractiveness is detected in all the 
aspects of the traffic contemplated: the rate of exploration of new sites by 
the users, the capture of new visitors by the sites or the new clicks 
received on each connection user-Web page. In all these cases, the linear
relation holds in short period of time. For longer periods, the 
life-time of 
the Web pages must be taken into account, complicating substantially the scenario. Preferential linking,
priority queuing and Web page aging seem thus to be essential factors for 
any model aimed to characterize Web surfing. 

{\it Acknowledgments---} The authors would like to thank Alain Barrat, Stefan Boettcher, Ciro 
Cattuto, Helmut Katzgraber, Filippo Menczer, Muhittin Mungan, Filippo Radicchi, and in general 
the members of the Cx-Nets collaboration for useful 
discussions and comments. We would also like to thank the IT service of Emory 
University for access to the database. Funding from the 
Lagrange Project of the CRT 
Foundation (Torino, Italy) and from the National Science Foundation under grant number 
$0312150$ was received. The use of computer resources provided by the 
Open Science Grid supported by the NSF and by the Office of Science 
of the U.S. Department of 
Energy is acknowledged.


\begin{thebibliography}{99}


\bibitem{berners-lee06}
T. Berners-Lee , W. Hall,J. Hendler , N. Shadbolt and D.J. Weitzner,
Science {\bf 313}, 769 (2006).

\bibitem{watts07}
D.J. Watts,
Nature {\bf 445}, 489 (2007).

\bibitem{romu04}
R. Pastor-Satorras and A. Vespignani,
{\it Evolution and Structure of the Internet : A Statistical Physics
  Approach}, Cambridge University Press (2004).

\bibitem{sergei03}
S. Dorogovtsev and J.F.F. Mendes,
{\it Evolution of Networks: From Biological nets to the Internet and
  WWW}, Oxford University Press (2003).

\bibitem{watts98}
D.J. Watts and S.H. Strogatz,
Nature {\bf 393}, 409 (1998).

\bibitem{huberman98}
B.A. Huberman, P.L. Pirolli, J.E. Pitkow and R.M. Lukose,
Science {\bf 280}, 95 (1998).


\bibitem{barabasi99}
A.-L. Barab\'asi and R. Albert,
Science {\bf 286}, 509 (1999).

\bibitem{menczer04}
F. Menczer (2004),
Proc. Nat. Acad. Sci. {\bf 99}, 14014 (2004).

\bibitem{sergei01}
S.N. Dorogovtsev and J.F.F. Mendes,
Phys. Rev. E {\bf 63}, 056125 (2001).

\bibitem{ciro06}
C. Cattuto, V. Loreto and V.D.P. Servedio,
Europhys. Lett. {\bf 76}, 208 (2006).

\bibitem{simkin07}
M.V. Simkin and V.P. Roychowdhury,
EuroPhys. Lett. {\bf 82}, 28006 (2007).

\bibitem{wu07}
F. Wu and B.A. Huberman,
Proc. Nat. Acad. Sci. {\bf 104}, 17599 (2007).

\bibitem{borgatta00}
E.F. Borgatta and R.J.V. Montgomery (editors),
{\it Encyclopedia Of Sociology - Volume I}, Macmillan Reference USA, 
2nd edition (2000).

\bibitem{onnela07}
J.-P. Onnela {\it et al.},
Proc. Nat. Acad. Sci. {\bf 104}, 7332 (2007).

\bibitem{golder06}
S. Golder, D. Wilkinson and B.A. Huberman,
{\tt  e-print ArXiv cs/0611137} (2006).


\bibitem{candia07}
J. Candia {\it et al.}, 
{\tt e-print ArXiv cond-mat/0710.2939} (2007).

\bibitem{vazquez07} 
A. V\'azquez,
Physica A {\bf 373}, 747 (2007).



\bibitem{meiss08} 
M.R. Meiss, F. Menczer, S. Fortunato, A. Flammini and A. Vespignani A,
{\it Ranking Web sites with real user traffic},
Proc. WSDM (2008).

\bibitem{zhou1}
T. Zhou, X.-P. Han, and B.-H. Wang, 
{\tt arXiv: 0801.1389} (2008).

\bibitem{zhou2}
T. Zhou, H.-A.T. Kiet, B.J. Kim, B.-H. Wang and P. Holme, 
EuroPhys. Lett. {\bf 82}, 28002 (2008).


\bibitem{albert99}
R. Albert, H. Jeong and A.-L. Barab\'asi,
Nature {\bf 401}, 130 (1999).

\bibitem{dill02} 
S. Dill {\it et al.},
\newblock {ACM Transactions on Internet Technology} {\bf 2}, 205 (2002).

\bibitem{meiss05} 
M.R. Meiss, F. Menczer and A. Vespignani,
{\it On the lack of typical behavior in the global Web traffic network},
Proc. WWW (2005).

\bibitem{catledge95}
L.D. Catledge and J.E. Pitkow,
Computer Networks and ISDN Systems {\bf 27}, 1065 (1995).

\bibitem{ciro07}
C. Cattuto, V. Loreto and L. Pietronero,
Proc. Nat. Acad. Sci. {\bf 104}, 1461 (2007).

\bibitem{capocci06}
A. Capocci {\it et al.}, 
{\tt e-print ArXiv physics/0602026} (2006).

\bibitem{vazquez06} 
A. V\'azquez {\it et al},
Phys. Rev. E {\bf 73}, 036127 (2006).

\bibitem{barabasi05}
A.-L. Barab\'asi,
Nature {\bf 435}, 207 (2005).

\bibitem{oliveira05} 
J.G. Oliveira and A.-L. Barab\'asi,
Nature  {\bf 437}, 1251 (2005).
 
\bibitem{vazquez05} 
A. V\'azquez,  
Phys. Rev. Lett. {\bf 95} 248701 (2005).

\bibitem{oliveira07} 
J.G. Oliveira and A. V\'azquez,
{\tt e-print ArXiv 0710.4916} (2007).

\bibitem{simon55} 
E.H. Simon,
Biometrika {\bf 42}, 425 (1955).

\bibitem{bornholdt01} 
S. Bornholdt and H. Ebel,
Phys. Rev. E {\bf 64}, 035104(R) (2001).

\bibitem{dorogovtsev00}
S. Dorogovtsev, J.F.F. Mendes and A.N. Samukhin,
{\tt e-print ArXiv condmat 0009090} (2000).

\bibitem{barabasi02}
A.-L. Barab\'asi A-L {\it et al},
Physica A {\bf 311}, 590 (2002).

\bibitem{romu01}
R. Pastor-Satorras, A. V\'azquez and A. Vespignani, 
Phys. Rev. Lett.  {\bf 87}, 258701 (2001).


\bibitem{redner05}
S. Redner, 
Physics Today {\bf 58}, 49 (2005). See e-print ArXiv physics/0506056 for a
more extense version (2005).


\bibitem{kraprivsky00}
P.L Krapivsky, S. Redner and F. Leyvraz,
Phys. Rev. Lett. {\bf 85}, 4629 (2000).

\bibitem{kraprivsky01}
P.L. Krapivsky and S. Redner,
Phys. Rev. E {\bf 63}, 066123 (2001).

\end{thebibliography}
\end{document}